\documentclass[prl,superscriptaddress,showpacs,longbibliography,reprint]{revtex4-2}
\usepackage[utf8]{inputenc}
\usepackage{amsfonts}
\usepackage{amsmath}
\usepackage{amssymb}
\usepackage{geometry}
\usepackage{braket}
\usepackage{dsfont}
\usepackage{color}
\usepackage[colorlinks = true, citecolor = blue, linkcolor = blue, urlcolor=blue]{hyperref}
\usepackage[normalem]{ulem}

\DeclareMathOperator{\Tr}{\operatorname{Tr}}

\newcommand{\traj}{\omega_0^T}

\newcommand{\condarg}{\left(S_t|S_{t-1}\right)}
\newcommand{\tranarg}{\left(S_t,S_{t-1}\right)}

\newcommand{\vfa}{v_{\psi}(S)}
\newcommand{\pfa}{\pi_{w}(a|S)}

\newcommand{\A}{\mathcal{A}}

\newcommand{\bias}{\lambda}

\usepackage{graphicx}
\graphicspath{{figures/}}

\usepackage{soul}

\begin{document}

\title{Combining Reinforcement Learning and Tensor Networks, with an Application to Dynamical Large Deviations}

\author{Edward Gillman}
\affiliation{School of Physics and Astronomy, University of Nottingham, Nottingham, NG7 2RD, UK}
\affiliation{Centre for the Mathematics and Theoretical Physics of Quantum Non-Equilibrium Systems,
University of Nottingham, Nottingham, NG7 2RD, UK}

\author{Dominic C. Rose}
\affiliation{Department of Physics and Astronomy, University College London, Gower Street, London WC1E 6BT, United Kingdom}

\author{Juan P. Garrahan}
\affiliation{School of Physics and Astronomy, University of Nottingham, Nottingham, NG7 2RD, UK}
\affiliation{Centre for the Mathematics and Theoretical Physics of Quantum Non-Equilibrium Systems,
University of Nottingham, Nottingham, NG7 2RD, UK}

\begin{abstract}
We present a framework to integrate tensor network (TN) methods with reinforcement learning (RL) for solving dynamical optimisation tasks. We consider the RL actor-critic method, a model-free approach for solving RL problems, and introduce TNs as the approximators for its policy and value functions. Our ``actor-critic with tensor networks'' (ACTeN) method is especially well suited to problems with large and factorisable state and action spaces. As an illustration of the applicability of ACTeN we solve the exponentially hard task of sampling rare trajectories in two paradigmatic stochastic models, the East model of glasses and the asymmetric simple exclusion process (ASEP), the latter being particularly challenging to other methods due to the absence of detailed balance. With substantial potential for further integration with the vast array of existing RL methods, the approach introduced here is promising both for applications in physics and to multi-agent RL problems more generally.
\end{abstract}

\maketitle
    
\noindent \textbf{\em Introduction.} 
Tensor networks (TNs), routinely used in the study of quantum many-body systems \cite{White1992, White1993, Schollwock2011, Eisert2010, Eisert2013, Montangero2018, Orus2019}, are increasing being applied in machine learning (ML), see e.g.\ Refs.~\cite{Stoudenmire2016,Liu2018,Sun2019,Roberts2019,Ghahramani1997,Efthymiou2019,Stoudenmire2018,Han2018,Gao2019,Levine2019,Guo2018,Glasser2018,Glasser2019}. Both domains often deal with systems with state-spaces which are exponentially large in the  number of degrees of freedom, say the number of qubits in a quantum system, or of pixels in images to be classified. In such situations TNs provide a powerful way to represent functions, vectors and distributions, while allowing for efficient sampling and computation of quantities such as inner products and norms.

To date, the intersection of TNs and ML has been mostly in supervised and unsupervised learning, see e.g.\ Refs.~\cite{Stoudenmire2016,Liu2018,Sun2019,Roberts2019,Ghahramani1997,Efthymiou2019,Stoudenmire2018,Han2018,Gao2019,Levine2019,Guo2018,Glasser2018,Glasser2019}. In contrast, the combination of TNs and reinforcement learning (RL) \cite{Sutton2018} 
has been more limited, despite recent major advances in RL \cite{Mnih2015,Haarnoja2018,Silver2016,Schulman2015,Wang2019}. While some promising related directions have been explored, such as the approximation of
Q-functions in the context of large state-spaces \cite{Metz2022}, and/or large action-spaces \cite{Mahajan2021}, the flexible integration of TNs with RL remains an open problem, along with demonstrating useful applications.

\begin{figure}[t!]
    \centering
    \includegraphics[width=0.72\linewidth]{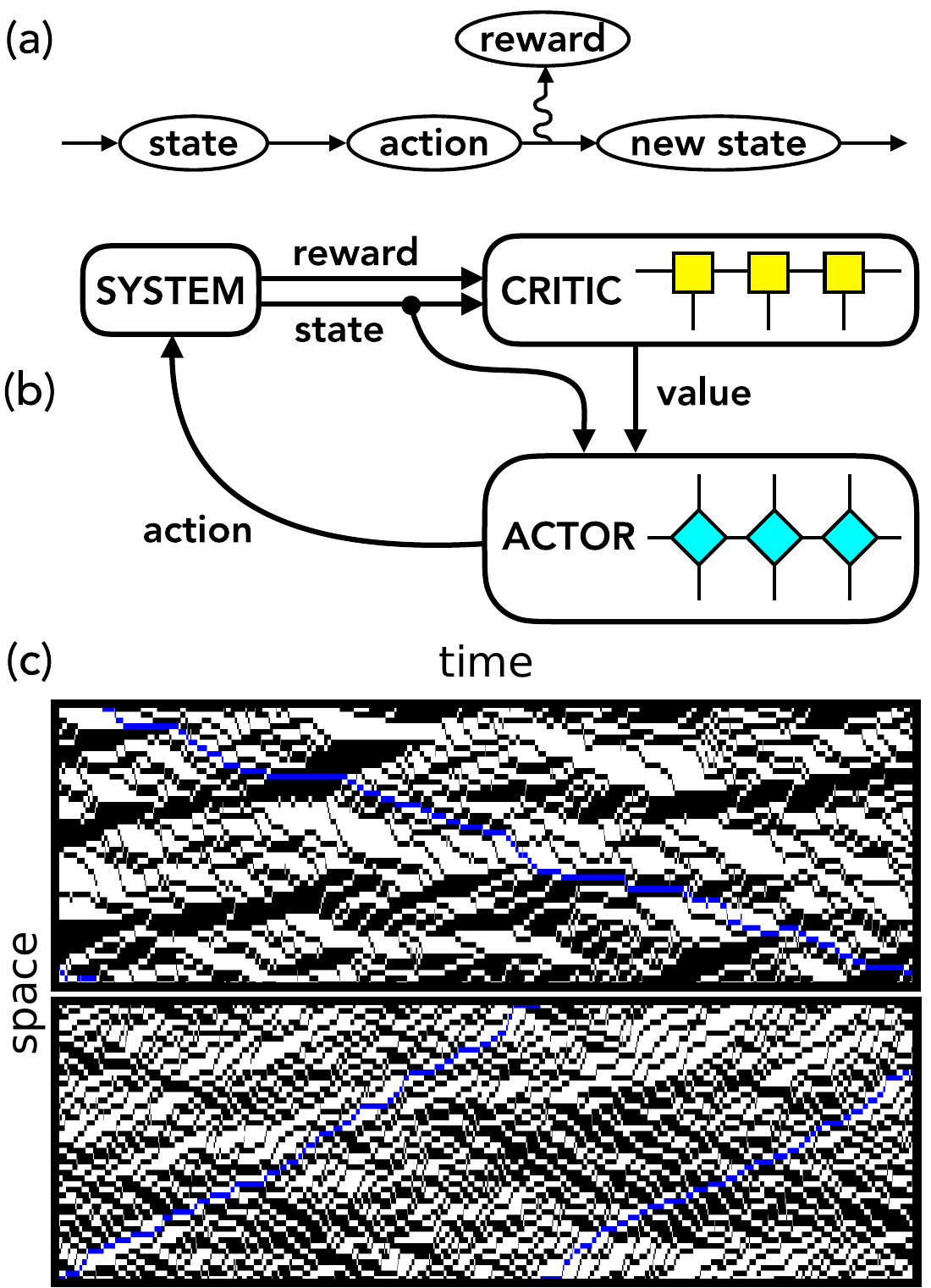}
    \caption{
    \textbf{Actor-Critic with tensor networks (ACTeN) } 
    \textbf{(a)} Sketch of a Markov decision process. 
    \textbf{(b)} In actor-critic RL, the state is passed to an ``actor'', which chooses the action, and to a ``critic'', which 
    values the state given the reward. This value is used to improve the actor's policy. In ACTeN, the function approximators for actor and critic are tensor networks. 
    \textbf{(c)} 
    Top: typical trajectory of the ASEP at half-filling and $L=50$ sites with one particle highlighted (blue), shown for $3000$ steps. Bottom: trajectory with a current large deviation, sampled from the ACTeN solution for biasing (counting) field $\lambda=-3$. See the text for details.
    } 
    \label{fig:tn_ac}
\end{figure}
    
Here, we introduce the {\em actor-critic with tensor networks} (ACTeN) method, a general framework for integrating TNs into RL via actor-critic (AC) techniques. By combining decision-making ``actors" with ``critics" that judge an actor's quality, AC methods are used in many state-of-the-art RL applications. Using TNs as the basis for modelling actors and critics within AC and RL represents a powerful combination to tackle problems with both large state and action spaces. 

To demonstrate the effectiveness of our approach, we consider the problem of computing the large deviation (LD) statistics of dynamical observables in classical stochastic systems \cite{Touchette2009,Garrahan2016,Garrahan2018,Jack2015a,Jack2019,Chetrite2015,Casert2021}, and of optimally sampling the associated rare long-time trajectories \cite{Majumdar2015,Chetrite2015b,Bolhuis2002,Giardina2006,Giardina2011,Cerou2007,Lecomte2007,Whitelam2020,DeBruyne2021a,DeBruyne2021b,DeBruyne2021c,DeBruyne2022,Pozzoli2022}. Such problems are of wide interest in statistical mechanics and can be phrased straightforwardly as an optimization problems that may be solved with RL \cite{Rose2021,Das2019,Das2021,Das2022} and similar techniques \cite{Borkar2003,Ferre2018,Jiawei2022,Jiawei2022b,Nemoto2014,Nemoto2017,Kappen2016,Bolhuis2022,Holdijk2022}. For concreteness we consider two models: (i) the East model, a kinetically constrained model used to study slow glassy dynamics, 
and (ii) the asymmetric exclusion process (ASEP), a paradigmatic model of non-equilibrium, in which particles hop around a lattice while blocking each others movement. In particular, and in contrast to the East model, the ASEP (with periodic boundaries) does not obey detailed balance 
\footnote{
    By obeying detailed balance we specifically mean that a system has a dynamical generator that can be made Hermitian via a similarity transformation defined through the stationary state (i.e.\ equilibrium) distribution. 
}, and thus evades straightforward use of TNs to compute spectral properties of the relevant dynamical generators. We demonstrate that ACTeN can be applied to both problems irrespective of the equilibrium/non-equilibrium distinction, by computing their dynamical LDs for sizes well beyond those achievable with exact methods. Given the vast array of options for improving the RL algorithm that we use, our results indicate that the overall framework outlined here is highly promising for applications more generally.

\smallskip 

\noindent \textbf{\em Background: Reinforcement Learning and Actor-Critic.} 
A discrete-time Markov decision process (MDP) \cite{Sutton2018} consists at each time $t \in [0, T]$ of stochastic variables $X_{t} = \left(S_{t}, a_{t}, R_{t} \right)$, named \textit{state}, \textit{action} and \textit{reward}. We assume these are drawn from $t$-independent finite sets, $\mathcal{S}, \mathcal{A}$ and $\mathcal{R}$, where the action set may depend on the current state, $\mathcal{A}(S)$ for $S\in\mathcal{S}$. The action and state variables are associated with the \textit{policy}, $\pi(a|S)$, and \textit{environment}, $P(S' | S, a)$, distributions. These are sampled in a sequence of steps to generate a trajectory of the MDP, $\omega = (X_{0}, X_{1}, ..., X_{T})$, see Figure~\ref{fig:tn_ac}(a). We assume that the reward is a deterministic function of the state and action variables. 

In the typical scenario of \textit{policy optimisation}, the policy is controllable and known, while the environment is fixed and potentially unknown. 
We focus on MDPs that are ``continuing'', and admit a steady state distribution independent of $X_0$. 
We can then define the average reward per time-step when following a given policy as $r(\pi) = \lim_{t \to \infty} \mathds{E}_{\pi}[R_{t}]$, where $\mathds{E}_{\pi}\left[\cdot\right]$ is the stationary state expectation over states and over transitions from those states according to policy $\pi$. 
The task of policy optimisation is to find the policy $\pi^{*}$ that maximises $r(\pi)$.

Reinforcement learning (RL) refers to the group of methods that aim to discover optimal policies by using the experience gained from sampling trajectories of an MDP.
In \textit{policy gradient} methods, the policy is approximated by a function $\pfa$ with parameters $w$, and optimized using the gradient of $r(\pi_{w})$ with respect to $w$. Building on this, actor-critic (AC) methods then assess $\pfa$ by computing the value, $v_\pi(S)$, of the states that result when following the policy. This value is defined as the difference between rewards in the future of a given state and the average reward, $v_\pi(S)=\mathds{E}_\pi \left[\sum_{\tau = t}^{\infty} \left[R_{\tau} - r(\pi)\right]|S\right]$.
The gradient of $r(\pi_w)$ can be written exactly in terms of these values as
\begin{align}
    \nabla_{w} r(\pi) &=  \mathds{E}_\pi \left[\delta^\pi_{t} \nabla_{w}\ln \pi_{w}(a_{t}|S_{t}) \right] ~,
\label{eqn:policy_gradient}
\end{align}
where we have introduced the temporal difference (TD) error, $\delta_{t}^\pi=v_{\pi}(S_{t+1}) + R_{t+1} - r(\pi_{w}) - v_{\pi}(S_{t})$ \cite{Sutton2018,Rose2021}, which quantifies if a resultant state is better than the current one. AC methods use this information to alter the probability of taking that action in the future.

In reality, calculating the true value of every state under the current policy is impractical, and thus an auxiliary approximation for the value-function, $\vfa$ with parameters $\psi$, is introduced, the so-called ``critic''.
To optimize the critic, we note that the value of a state is related to the value of states reachable from it, as encoded in the differential Bellman equation \cite{Sutton2018}, 
\begin{align}
    v_\pi(S_t)=\mathds{E}_\pi\left[v_\pi(S_{t+1})+R_{t+1}-r(\pi_w)|S_t\right].
    \label{eqn:differential-bellman}
\end{align}
Minimizing the error in the Bellman equation when substituting the critic for the true values can be done by updating the weight as $\psi'=\psi+\alpha\mathds{E}_\pi \left[\delta_{t} \nabla_{\psi}\vfa(S_{t}) \right]$, in terms of the approximate TD error, $\delta_{t}=v_\psi(S_{t+1}) + R_{t+1} - r(\pi_{w}) - v_\psi(S_{t})$, and a learning rate $\alpha$.
Intuitively, the estimated expected reward before a transition occurs, $v_\psi(S_{t})$, is compared to the expected reward afterwards plus the true reward for that time-step, $v_\psi(S_{t+1})+R_{t+1}-r(\pi_w)$, and $\vfa$ adjusted to make these closer. The policy is then updated by following the gradient in Eq.~\eqref{eqn:policy_gradient} with the exact TD error $\delta^\pi$ replaced by the critic's approximate TD error $\delta$.

\smallskip

\noindent 
\textbf{\em Analytic Example: Two-Site East Model.} 
To illustrate these ideas, consider two spins, $s_{1,2} = 0,1$, evolving with a constrained set of transitions as in the East model studied below, such that a spin can flip only when its left-neighbour (in this case simply the other spin due to periodic boundaries) is $1$. The states are
$\mathcal{S} = \{ 00,01,10,11 \}$. The dynamics of this model can be implemented as an MDP using a policy that (stochastically) selects which spins to flip. Denoting no-flip/flip by $0/1$ and requiring at most one spin flip per time-step, the action-sets for this model are then: $\mathcal{A}(10)=\{00,01\}$, $\mathcal{A}(01)=\{00,10\}$, and $\mathcal{A}(11)=\{10,01\}$, with the state $00$ being disconnected from the rest. An example policy that selects from all possible actions equally would assign a probability of $1/2$ to each of these, e.g. $\pi(a=00|S=10) = \pi(a=01|S=10) = 1/2$ and similarly for the other states. The transition to a new state is then enacted by the environment. We can choose this to be deterministic, and simply apply the spin-flip operations selected by the actor to the current state, $s_i \to (1 - a_i) s_i + a_{i} (1 - s_{i})$. Finally, an example reward can be defined via the function $R(S,a,S')=-\lambda(1-\delta_{S',S})$, which awards $-\lambda$ every time a spin flip occurs ($a\neq0$), encouraging activity/inactivity for negative/positive $\lambda$. For this reward, the optimal policy for negative $\lambda$ will maximize activity, flipping a spin at every step, requiring going from $10$ and $01$ to $11$ with probability $1$, i.e. $\pi^{*}(01|10) = 1$, etc.

\smallskip

\noindent 
\textbf{\em Method: Actor-Critic with Tensor Networks (ACTeN).} 
We now focus on applying AC to solve problems with large state and action spaces. For example, we may wish to find the optimal dynamics of a system of $L$ binary components, resulting in $2^L$ states, with individual agents and their actions associated to each component. To ensure optimal choices for a given task, actions may need to be correlated not only with other agents states, but also the actions other agents are about to take. In such problems, simple approaches such as tabular RL fail due to exponentially large memory requirements and sampling costs, and a common alternative is to use neural networks (NNs) when defining $\pfa$ and $\vfa$. TNs offer another approach, with polynomial memory and computational costs, and showing state-of-the-art performance in many settings; see e.g. the review \cite{Montangero2018}. Motivated by this, we define a general framework (which we call ACTeN) that exploits TNs to efficiently represent $\pfa$ and $\vfa$.

A TN is a set of tensors, $\mathcal{T} = \lbrace T_{i_{1}j_{1}k_{1}\cdots}^{[1]}, T_{i_{2}j_{2}k_{2}\cdots}^{[2]}, ... \rbrace$, contracted in some pattern, cf.\ Fig.~\ref{fig:tn_ac}(b). This results in a single tensor that can be viewed as defining a multivariate function, $\varphi(x)=T_x$, $T_{x} = \mathcal{C}[\mathcal{T}]$, where $\mathcal{C}$ indicates the chosen contractions and $x$ all remaining uncontracted indices. For a given problem, the selection of an appropriate TN  depends on factors such as dimensionality and geometry. 
Here we consider applying ACTeN to study one-dimensional (1D) systems with periodic boundaries (PBs) and $L$ components, such that a state is $S=(s_1, \cdots, s_L)$ with $s_i$ taking $d$ values. To represent the value function $\vfa$ we use a translation invariant matrix product state (MPS) which mirrors the chain geometry of the system. This TN is built from a single real-valued tensor $\A_{sij}$ of shape $(d,\chi,\chi)$ (or equivalently $d$ square $\chi$-dimensional matrices $[A_s]_{ij}=A_{sij}$) whose elements encode the parameters of $\vfa$, $\psi = \lbrace A_{sij} \rbrace_{s=1}^d$. For a given $S=(s_1, \cdots, s_L)$ this TN is defined as,  
\begin{align}
    \varphi(S) &= \Tr\left[A_{s_{1}} A_{s_{2}} ... A_{s_{L}}\right] ~,
\label{eqn:mps_ti_def}
\end{align}
i.e., for each site we select the corresponding matrix $A_{s_{i}}$, multiplying the $L$ matrices together with a trace to produce a real scalar. To take advantage of translation invariance and apply approximations from smaller $L$ to larger $L$ systems, we then define the value function in terms of $\varphi(S)$ after the additional application of a square and log, which prevents the exponential growth or decay of values as $L$ is changed for fixed $A_{sij}$. Hence,
\begin{align}
    \vfa &= \log\left[\varphi(S)^2\right] ~.
\label{eqn:vf_defn}
\end{align}

To define $\pfa$ we use a matrix product operator (MPO). This TN is built from a single real-valued $(d_S,d_A,\chi,\chi)$-shaped tensor, $\mathcal{A}_{asij}$, equivalent to $d_s\times d_A$ $\chi$-dimensional square matrices $[A_{as}]_{ij}=A_{asij}$, i.e., one matrix per combination of local state and action. Given a state $S=(s_1, \cdots, s_L)$ and actions $a=(a_1, \cdots, a_L)$, the contraction is given by the traced matrix product,
\begin{align}
    \varphi(a, s) &= \Tr\left[A_{a_{1}s_{1}} A_{a_{2}s_{2}} ... A_{a_{L}s_{L}}\right] ~.
\label{eqn:mpo_ti_def}
\end{align}
To use this to define a policy, we need to ensure positivity and normalization, as well as preventing the policy from producing invalid actions. To achieve this we define
\begin{align}
\pfa = \mathcal{C}(a, S)[\mathcal{N}(S)]^{-1}\varphi(a,S)^{2} ~ ,
\label{eqn:policy_defn}
\end{align}
where $\mathcal{N}(S) = \sum_{a | \mathcal{C}(a, S) = 1} \varphi(a,S)^{2}$ is the (state-dependent) normalisation factor and $\mathcal{C}(a, S)$ returns one if an action $a$ is possible in state $S$ or zero otherwise
\footnote{There are two situations where one might simply set $\mathcal{C}(a, S)=1$ for all state-action pairs: (i) when all actions are possible from any state; (ii) the form of the constraint is unknown and must be learnt, for example by penalising the policy when disallowed actions are selected.}.

\begin{figure}[t]
    \centering
    \includegraphics[width=1.0\linewidth]{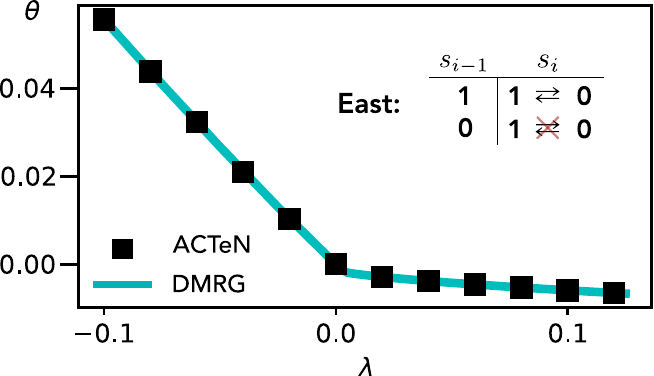}
    \caption{\textbf{Dynamical large deviations in the East model using ACTeN.} Scaled-cumulant generating function for the dynamical activity of the East model as a function of biasing field $\lambda$ from ACTeN (symbols), for $L=50$ and PBC. 
    Our RL results coincide with those obtained from the current state-of-the-art method using DMRG, cf.\ Ref.~\cite{Banuls2019} (which is possible since the East model obeys detailed balance). Inset: Kinetic constraint of the East model; a spin, $s_{i}$, can flip, $s_{i} \to 1 - s_{i}$, only if the spin to the left is up, $s_{i-1} = 1$.}
    \label{fig:em}
\end{figure}

\begin{figure*}[t]
\centering
\includegraphics[width=1\linewidth]{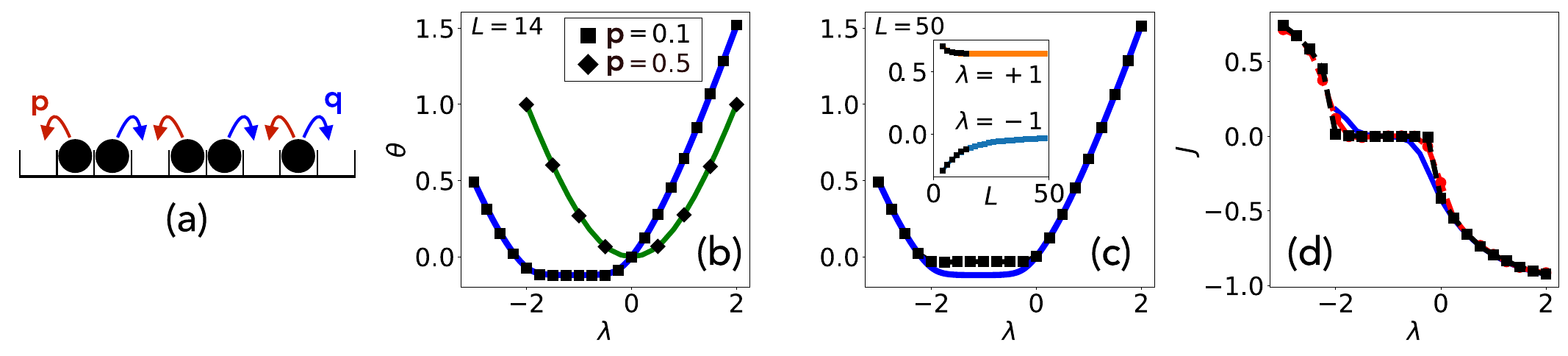}

\caption{\textbf{Dynamical large deviations in the ASEP using ACTeN.} \textbf{(a)} In the ASEP particles can only move to an unoccupied neighbouring site, with probability $p$ to the left and $q=1-p$ to the right.  \textbf{(b)} SCGF for the time-integrated particle current
as a function of biasing field. We show results from ACTeN for $p=0.1$ (squares) and $p=1/2$ (diamonds). The lack of detailed balance for PBC and $p \neq 1/2$ prevents straightforward application of DRMG, but for small sizes (here $L=14$) we can compare to exact diagonalisation (blue curve for $p=0.1$, green for $p=1/2$). 
\textbf{(c)} SCGF for $p=0.1$ from ACTeN for size $L=50$ which is beyond the scope of ED. Compared to $L=14$ (blue curve from ED), we see that ACTeN captures the flattening of the SCGF for larger sizes indicative of a LD phase transition, cf.\ Ref.~\cite{Jack2015a}. The inset shows the smooth convergence of our ACTeN numerics with $L$ for two values of $\lambda$. \textbf{(d)} Since ACTeN provides direct access to the optimal dynamics, observables such as the time-integrated current can be evaluated directly (black squares for $L=50$). We show for comparison the numerical differentiation of the ACTeN SCGF (red circles) and of the ED SCGF at $L=14$ (blue line). 
}
\label{fig:asep}
\end{figure*}

\smallskip

\noindent \textbf{\em Application: Dynamical Large Deviations.} 
To test ACTeN we consider the problem of computing the large deviations (LDs) of trajectory observables 
\cite{Touchette2009,Garrahan2018,Jack2019} in the East model and the ASEP in 1D with PBs. Both models are many-body binary spin systems with large state-spaces for large $L$, $S = \{s_i\}_{i=1}^L$, with $s_{i}=0, 1$ (see the Appendix for details on the models). The dynamics of these systems are subject to local constraints that lead to rich behaviours in their trajectories, $\traj = \{S_{t}\}_{0}^{T}$. This can be observed in the time-integrals of time-local quantities, $O(\traj) = \sum_{t=1}^{T} o(S_{t}, S_{t-1})$, the moments of which are contained in derivatives of the moment generating function (MGF), $Z_T(\lambda) = \sum_{\traj} e^{-\lambda O(\traj)}P(\traj)$, where $P(\traj)=\prod_{t=1}^TP(S_t|S_{t-1})P(S_0)$ is the trajectory probability under the dynamics.

In the long-time limit, the MGF obeys a large deviation principle \cite{Touchette2009,Garrahan2018,Jack2019} 
with the scaled cumulant generating function (SCGF), $\theta(\lambda) = \lim_{T \to \infty} \frac{1}{T}\ln Z_T(\lambda)$, playing the role of a free-energy for trajectories. In principle the SCGF can be obtained by sampling methods. 
However, this is exponentially hard (in time and space) using the original dynamics. An alternative is to find a more efficient sampling dynamics which may then be combined with importance sampling to obtain unbiased statistics.
This can be formulated as a RL problem as follows: can we find a parameterized dynamics $P_{w}\condarg$ such that $P_w(\traj)=e^{-\lambda O(\traj)}P(\traj)/Z_T(\lambda)$, i.e. it reproduces a trajectory ensemble biased towards rare trajectories of the original dynamics.
This dynamics is connected to an underlying policy $\pi_{w}(a|S)$ by a deterministic environment which returns states after receiving an associated action, i.e. $P_w\left[S'=f(a,S)|S\right]=\pi_w(a|S)$, where for each $S$, $f(a,S)$ returns a unique $S'$ for each $a$. For example, in the East model if we take the action $a=\{a_i\}_{i=1}^L$ then sites with $a_{i} = 1$ are flipped and those with $a_{i}=0$ are not flipped; the new state is then $S' = f(a,S) = \lbrace(1-a_{i})s_{i} + a_{i}(1-s_{i}) \rbrace_{i=1}^{L}$.

Optimizing the KL divergence between the two trajectory ensembles gives a regularized form of RL with a reward depending on the policy \cite{Rose2021}
\begin{align}
    R_{t}= - \lambda o(S_{t}, S_{t-1})- \ln\left(\frac{P_w\condarg}{P_{\rm orig}\condarg}\right),
\label{eqn:biased_dynamics_return}
\end{align}
with its expected value becoming the SCGF at optimality \cite{Rose2021}.
Intuitively, choosing actions (e.g. flips) to maximize the first term increases the likelihood of rare events with extreme values of the observable, while maximizing the second term minimizes the difference between the parameterized and original dynamics, thus making the event more probable.
Maximizing this reward is a balancing act between these two aims, resulting in dynamics biased towards rare events in a way representative of their occurrence in the original dynamics.
In the appendices, we illustrate ACTeN by solving explicitly the 2-site East model and showing how this can be exactly represented by the TN ansatz.

\smallskip 

{\em (i) East model and dynamical activity:} Figure \ref{fig:em} shows the SCGF of the dynamical activity [total number of spin flips in a trajectory, defined by $o\tranarg = 1- \delta_{S_{t},S_{t-1}}$], calculated using ACTeN (symbols). 
Since the East model obeys detailed balance, the SCGF is the log of the largest eigenvalue of a Hermitian operator and can be estimated via density matrix renormalisation group (DMRG) methods, cf.\ Ref.~\cite{Helms2019,Banuls2019,Helms2020,Causer2020,Causer2021,Causer2022,Causer2022b} (here we \texttt{ITensors.jl} \cite{Fishman2022}). 
Figure \ref{fig:em} shows that the DMRG results (blue curve) coincide with ACTeN (black squares) for size $L=50$, which is well beyond what is accessible to exact diagonalisation (ED). 
Note that DMRG with PBs tends to be much less numerically stable than for open boundaries. 
Nonetheless, ACTeN can reach $L \gtrsim 50$ without the need for any special stabilisation techniques. 

\smallskip 

{\em (ii) ASEP and particle current:} Figure \ref{fig:asep} presents the LDs of the time-integrated particle current, defined by $o(s_t, s_{t-1}) = \frac{1}{2}\sum_{i=1}^L s_{t-1}^{i}s_{t}^{i+1} - s_t^is_{t-1}^{i+1}$. Figure \ref{fig:asep}(b) shows the SCGF obtained via ACTeN (black squares/diamonds). 
Unlike the East model, for asymmetric hops ($p \neq 1/2$) Hermitian DMRG cannot be applied directly to the ASEP, so for comparison we show results from exact diagonalisation for both $p=0.1$ (blue line) and $p=1/2$ for $L=14$. 
Beyond $L=14$ ED becomes prohibitive, while ACTeN remains feasible. 
Figures~\ref{fig:asep}(c,d), show the expected phase transition behaviour \cite{Jack2015a} and convergence with $L$ up to $L=50$. The optimal dynamics itself, i.e. the learnt policy, can be used to generate trajectories representative of $\lambda \neq 0$, see Fig.~\ref{fig:tn_ac}(c), and directly sample rare values of the integrated current, see Fig.~\ref{fig:asep}(d).

\smallskip 

\noindent \textbf{\em Outlook.} ACTeN compares very favourably with state-of-art methods for computing rare events without some of the limitations, such as boundary conditions or detailed balance. From the corpus of research in both
TNs and RL, our approach has considerable potential for further improvement and exploration. These include: numerical improvements to precision via hyper-parameter searches; stabilisation strategies for large systems; integration with trajectory methods such as transition path sampling or cloning; integration with advanced RL methods such as those offered by the DeepMind ecosystem \cite{DeepMindJAX}; generalisation to continuous-time dynamics; and applications to other multi-agent RL problems, such as PistonBall \cite{terry2020pettingzoo}, via integration with additional processing layers particularly those for image recognition.

\smallskip 

\noindent \textbf{Acknowledgement.} We would like to thank Christopher J. Turner for useful discussions. We acknowledge funding from The Leverhulme Trust grant no.\ RPG-2018-181, EPSRC Grant No. EP/V031201/1, and University of Nottingham grant no.\ FiF1/3. We are grateful for access to the University of Nottingham's Augusta HPC service. DCR was supported by funding from the European Research Council (ERC) under the European Union’s Horizon 2020 research and innovation programme (Grant agreement No. 853368). We thank the creators and community of the Python programming language \cite{VanRossum2009}, and acknowledge use of the packages \texttt{JAX} \cite{Jax2018}, \texttt{NumPy} \cite{Harris2020}, \texttt{pandas} \cite{Pandas2022}, \texttt{Matplotlib} \cite{Hunter2007} and \texttt{h5py} \cite{Collette2014}. We also thank the creators and community of the Julia programming language \cite{Bezanson2017}, and acknowledge use of the packages \texttt{ITensors.jl} \cite{Fishman2022} and \texttt{HDF5.jl} \cite{HDF5jl}.

\smallskip 

\noindent \textbf{Appendix on kinetically constrained models:}

\textbf{\em (i) East Model.}
Flipping of a spin is constrained on the spin to its left being in state $1$ \cite{Jackle1991,Garrahan2018}.
Dynamics then amounts to two steps: first, select a random site $i$ with probability $1/L$; second, if spin $s_{i-1}=1$ then flip $s_{i}$. Given $N$ spins in state $1$ and periodic boundary conditions, the transition probability is $P(s'|s)=\frac{1}{L}$ for each possible new state $s'\neq s$, and probability $P(s|s)=1-\frac{N}{L}$ for no flip occurring.

\textbf{\em (ii) Asymmetric simple exclusion process.}
The constraint is particle exclusion: a particle at site $i$ ($s_i=1$) can move left or right only if the destination is unoccupied \cite{Spitzer1970}. The movement of a particle to, say, the right thus corresponds to $10 \to 01$, i.e. a flip of both spin variables. The dynamics again amounts to two steps: first, select a particle with probability $1/N$, with $N=\sum_is_i$ the particle number; second, choose whether this particle hops right or left with probabilities $p$ or $1-p$, respectively, with the hop occurring if the new site is unoccupied.
The transition probabilities are then: $P(s'|s)=p/N$ for a right hop; $P(s'|s)=(1-p)/N$ for a left hop; and, given the number of neighbouring particles $N_\textrm{nn}=\sum_{i=1}^Ls_is_{i+1}$, with $s_{L+1}:=s_1$, the probability of no change is $P(s|s)=1-N_\textrm{nn}/N$. In the main text, we consider the case of half-filling, $N=L/2$.

\smallskip 

\noindent \textbf{Appendix on analytical solution of the two-site East model with tensor network ansatz:} 

An exact solution for the optimal policy in the two-site East model can be found by analytically constructing the so-called ``Doob dynamics'' \cite{Jack2010,Chetrite2015} as follows. First we define a ``tilted'' evolution operator \cite{Touchette2009,Garrahan2018,Jack2019} $P_\lambda\condarg=e^{-\lambda o(S_{t}, S_{t-1})}P\condarg$, such that $Z_T(\lambda) = \left\langle -\right|P_\lambda^T\left|P_{\rm ss}\right\rangle$ where $\left\langle -\right|=\sum_S\left\langle S\right|$ and $\left|P_{\rm ss}\right\rangle$ is the stationary state vector for $P$.
In the large $T$ limit the SCGF is the log of the largest eigenvalue of $P_\lambda$ \cite{Touchette2009,Garrahan2018,Jack2019}.
The corresponding left eigenvector $l_\lambda$ is related to the dynamics that maximises the expected value of Eq.~\eqref{eqn:biased_dynamics_return}, given by the so-called Doob (or optimal) dynamics $P_\lambda^D\condarg=\frac{l_\lambda(S_{t-1})P_\lambda\condarg}{e^{\theta(\lambda)}l_\lambda(S_{t-1})}$.
Applied to the two-site East model, defining the function $a(\lambda)=\frac{4}{1+\sqrt{1+8e^{-2\lambda}}}$, we find $\theta(\lambda) = -\ln\left(a(\lambda)\right)$, with optimal dynamics 
\begin{align}
    P_\lambda^D(10|10)&=P_\lambda^D(01|01)=\frac{a(\lambda)}{2}, \nonumber\\
    P_\lambda^D(11|10)&=P_\lambda^D(11|01)= 1-\frac{a(\lambda)}{2}, \nonumber\\ P_\lambda^D(10|11)&=P_\lambda^D(01|11)=\frac{1}{2}. \nonumber
\end{align}

We may then find the corresponding value function by solving the differential Bellman equation \eqref{eqn:differential-bellman}. To do this, note first that by symmetry the values of states $10$ and $01$ are identical, and second that Eq.~\eqref{eqn:differential-bellman} is invariant under an overall shift of the value function by a constant. Therefore, we may choose the value of states $10$ and $01$ to be $0$, and thus find $V_\lambda(11)=-\lambda-\theta(\lambda)$.

These results are what is expected intuitively. Trajectories biased towards enhanced activity ($\lambda<0$) have $a(\lambda)<1$, making $P_\lambda^D(11|10)=P_\lambda^D(11|01)>1/2$, i.e. the system is more likely to transition to the state which is guaranteed to flip at the next step rather than remain in $01$ or $10$. Furthermore, $V_\lambda(11)>0$, i.e. the state guaranteed to flip is more valuable. In contrast, trajectories biased towards reduced activity ($\lambda>0$) show the opposite behaviour.

To connect $P_\lambda^D$ to our policy ansatz Eq.~\eqref{eqn:policy_defn}, we first rewrite it as an operator.
Using projection operators $P_1=\left|1\right\rangle\left\langle1\right|$, $P_0=\left|0\right\rangle\left\langle0\right|$, and flip operators $\sigma_+=\left|1\right\rangle\left\langle0\right|$, $\sigma_-=\left|0\right\rangle\left\langle1\right|$, we define $A(\lambda)=\sigma_- + a(\lambda)\sigma_+ + [2-a(\lambda)]P_0$.
We may thus write $P_\lambda^D=0.5(P_1 \otimes A(\lambda) + A(\lambda) \otimes P_1)$.

\begin{figure*}
\centering
\includegraphics[width=1\linewidth]{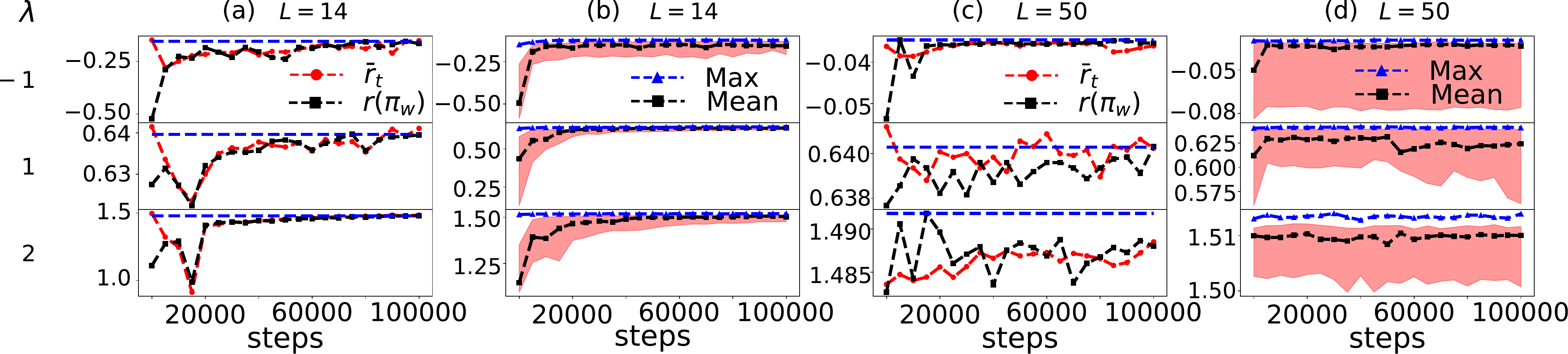}
\caption{
    \textbf{Training Procedure and Learning Curves (ASEP)} \textbf{(a)} For each bias [we show $\bias = -1$ (top row), $\bias =1$ (middle row), $\bias =2$ (bottom row)] TN-based policies and value-functions are produced via actor-critic optimization. 
    These are initiated at random for $L=4$ with $\chi=16$ and trained for $10^{6}$ steps. 
    Every $5000$ training steps the average reward of the policy is evaluated over $10^{4}$ steps (black squares) and the weights of the policy (which we call a ``snapshot'' for that time) are stored. 
    The evaluated values can be compared to the training estimate of $r(\pi)$ (red circles), which tends to overestimate $r(\pi)$ initially. 
    The policy snapshot with the highest evaluated $r$ (blue dashed line) is used to initiate the policy for higher values of $L$. 
    This is repeated every $\Delta L = 2$ up to $L=50$, with $L=14$ shown here. 
    \textbf{(b)} For each bias, several policies (here six) are independently trained via the same procedure from different random initial conditions. 
    This produces a distribution of evaluated average rewards, here represented by the median (black squares) and inter-quartile range (red-shaded region). 
    The policy with the maximum average reward at each $L$ is selected as the optimal dynamics (blue triangles). 
    \textbf{(c)} Same as (a) for $L=50$. The learning curves appear nosier than in (a) but note that the vertical scale is much smaller. The learning rate is kept fixed throughout. 
    \textbf{(d)} The distribution of $r$ across parallel agents for $L=50$ is again much tighter than for $L=14$.}
\label{fig_sm:asep_training}
\end{figure*}

The policy ansatz Eq.~\eqref{eqn:policy_defn} involves the element-wise square of an operator $\varphi(S',S)$: we thus seek the element-wise square root of $P_\lambda^D$.
We define $[\tilde{A}(\lambda)]_{ij}=\sqrt{[A(\lambda)]_{ij}}$.
Due to their sparsity structures, we have $\tilde{A}(\lambda)\odot P_1=0$, where $\odot$ is element-wise (Hadamard) multiplication.
Since $(A\otimes B) \odot (C \otimes D)$ = $(A\odot C)\otimes(B\odot D)$, we find $\varphi=\sqrt{0.5}(P_1 \otimes \tilde{A}(\lambda) + \tilde{A}(\lambda) \otimes P_1)$ is such that $\varphi\odot\varphi=P_\lambda^D$.
It remains to factor $\varphi$ into an MPO of the form of Eq.~\eqref{eqn:mpo_ti_def}. We may rewrite this as $\varphi=\sum_{d_1,d_2=1}^\chi T_{d_1d_2}\otimes T_{d_2d_1}$ where each $T_{d'd}$ is a $2\times2$ matrix acting on the single site state space.
This $T$ can be constructed by taking $\chi=2$ with $T_{11}=T_{22}=0$, $T_{12}={0.5}^{1/4}P_1$, and $T_{21}={0.5}^{1/4}\tilde{A}(\lambda)$.
We thus find this order-4 tensor $T$ reproduces the exact Doob dynamics from our translation invariant MPO-based ansatz.

For the value function, taking an exponential and square-root element-wise of the value function to invert Eq.~\eqref{eqn:vf_defn} leads to the vector 
$\tilde{V}^\lambda = \left(e^{V_\lambda(11)/2}-1\right)\ket{1}\otimes\ket{1}+\ket{-}\otimes\ket{-}$.
Since this is already a sum of symmetric products, it is easy to rewrite it as a translation invariant MPS with $\chi=2$, i.e. $\tilde{V}^\lambda_{s_1s_2}=\sum_{d_1,d_2=1}^2v_{s_1}^{d_1d_2}v_{s_2}^{d_2d_1}$, with order-3 tensor $v$ such that $v_1^{11}=\sqrt{e^{V_\lambda(11)/2}-1}$, $v_1^{22}=v_0^{22}=1$, and $v_{s}^{d'd}=0$ otherwise.

\smallskip

\noindent \textbf{Appendix on training procedure:}

We now provide more details on the training procedure used to obtain the policies of the main text. 
First, we outline the update step used to improve the policy and value function approximations.
Second, we outline size-annealing, where we apply transfer learning by using systems of increasing size. 
Finally, we discuss policy evaluation and selection, whereby the best policy is chosen from a set of candidates.

\textbf{\em (i) Basic Outline of Training.}
We start by initializing the parameters $w_{0}$, $\psi_{0}$ and $\bar{r}_{0}$, where $\bar{r}_{t}$ is an estimate of the average reward per-time step,  $r(\pi_{w})$, after $t$ training steps, along with the environment and initial state $s_{0}$. Choosing the three learning rates $\alpha_{\pi}, \alpha_{v}$ and $\alpha_{r}$, for each step $t \in [0, T]$ we:
\begin{enumerate}
    \item Sample an action $a_{t}\sim\pi_{w}( \cdot | s_{t})$ [where $x\sim Y(\cdot)$ stands for $x$ sampled from $Y$], and from it get its log probability and eligibility, 
    $\ln \pi_{w}(a_{t}|s_{t})$, $\nabla_{w} \ln\pi_{w}(a_{t}|s_{t})$.

    \item Get the next state and reward given the current state and action, $(s_{t+1}, r_{t+1}) \sim P(\cdot, \cdot |a_{t},s_{t})$.

    \item Get the temporal difference error with the current value function, $\delta_{t+1} = v_{\psi}(s_{t+1}) + r_{t+1} - \bar{r}_{t} - v_{\psi}(s_{t})$.
    
    \item Update the parameters of the value function, $\psi_{t+1} = \psi_{t} + \alpha_{v} \delta_{t+1} \nabla_{\psi} v_{\psi}(s_{t})$.
    
    \item Update the parameters of the policy, $w_{t+1} = w_{t} + \alpha_{\pi} \delta_{t+1} \nabla_{w} \ln\pi_{w}(a_{t}|s_{t})$ .
    \item Update the estimate of the average reward per time step, $\bar{r}_{t+1} = \bar{r}_{t} + \alpha_{r} \delta_{t+1}$ ~.
\end{enumerate}

\textbf{\em (ii) Annealing and Transfer Learning.}
In the context of machine learning, ``annealing'' (sequentially solving an optimisation problem reusing solutions to improve an initial guess) can be considered a form of {\em transfer learning}. In our case, we anneal the size of the system: the optimal policies for two system sizes will be similar as long as $L' \gtrsim L$, and in the settings considered the optimal dynamics should converge as $L \to \infty$.

We first approximate the optimal policy for a small system, $L=4$, starting from random initial weights. 
The weights after optimisation at this size are then used as the initial weights for $L = 6$. This is repeated in steps of $\Delta L = 2$, up to the maximum desired $L$. This process ensures that effectively much longer training times are used for larger systems, and produces smooth convergence curves in $L$, which can be used both as diagnostic tools and for extrapolation [c.f. Fig.~\ref{fig:asep}\textcolor{blue}{(c)} inset].

\textbf{\em (iii) Policy Evaluation and Selection.}
To determine the quality of a policy, we use it to generate trajectories without any change to the policy weights [c.f. Fig 1.(c) of the main text]. 
The set of rewards along these trajectories can then be averaged to estimate $r(\pi_{w})$ for the policy, allowing for different policies to be compared.

To ensure that we obtain the best policies possible, we then employ policy selection in two ways. 
Firstly, throughout training a given policy we store its weights periodically. 
After some number of periods, these weight snapshots are then evaluated and the best one is selected, ensuring that the policy can only improve with more training. 
Secondly, we run parallel policy optimisations and evaluations, starting from different random initial weights, with the best one selected.

The specific processes of policy evaluation and selection used to produce the results in the main text are illustrated for the ASEP in Fig. \ref{fig_sm:asep_training} (details in caption).

\bibliographystyle{apsrev}
\bibliography{TN_and_RL_bib}

\onecolumngrid
\widetext

\begin{center}
\textbf{\large Supplemental Material}
\end{center}

\setcounter{section}{0}
\setcounter{equation}{0}
\setcounter{figure}{0}
\setcounter{table}{0}
\setcounter{page}{1}
\makeatletter

\renewcommand\thesection{S\arabic{section}}
\renewcommand{\theequation}{S\arabic{equation}}
\renewcommand{\thefigure}{S\arabic{figure}}
\renewcommand{\thetable}{S\arabic{table}}
\renewcommand{\bibnumfmt}[1]{[S#1]}

\section{Forward Passes for the East Model and Simple Exclusion Process}

In this section we discuss the so-called forward passes required for the discovery of optimal dynamics using the function approximations described in the main text. Along with the information here, example scripts that implement the forward passes can be found on the associated GitHub repository, which should be referred to for more details \cite{SM_Code}. Note that, while we focus here on the case of kinetically constrained spin systems, specifically the east model and asymmetric simple exclusion process (ASEP), much of the following is generic and can easily be adapted to a variety of other applications.

Generally, to solve policy optimisation problems with actor-critic (AC), for a given state $s$ we must be able to:
\begin{enumerate}
    \item Evaluate the function approximation for the state-value function, $\vfa$.
    \item Sample an action, $a$, from the function approximation for the policy, $\pfa$.
    \item Calculate $\log\left[\pfa\right]$ for that action.
\end{enumerate}
The computations that implement these are known as the forward passes, while the so-called backward passes implement the gradients of these quantities with respect to the parameters (weights). Since gradients can be calculated automatically in programming frameworks such as \texttt{JAX} \cite{Jax2018}, we are only required to implement the necessary forward passes explicitly for a working AC method.

\subsection{Evaluation of $\vfa$}

Recall that in the main text we defined,
\begin{align}
    \vfa &= \log\left[\varphi(s)^{2}\right] ~,
\label{eqn_sm:vf_defn}
\end{align}
where
\begin{align}
    \varphi(s) &= \Tr\left[A_{s_{1}} A_{s_{2}} ... A_{s_{L}}\right] ~ .
\label{eqn_sm:mps_ti_def}
\end{align}

Since the form of $\vfa$ is independent of the dynamics in question, we start by describing its implementation here, before moving onto the specific implementation of the policy forward passes for each model in turn below. To ensure the implementation presented is readily applicable to automatic differentiation in frameworks such as \texttt{JAX} \cite{Jax2018}, we describe it here in terms of standard functions (specifically ``scan" functions) that are amenable to automatic differentiation.

For a set of steps, $k = 1,2,...,T$, a scan-function is defined by the repeated application of a given function, $f$, to a set of inputs, $x_{k}$, and a single ``carry" object, $C$, that is updated at each step. That is, for each $k$ the scan-function computes,
\begin{align}
    C, y_{k} = f\left(C, x_{k}\right) ~,
\end{align}
where $y_k$ is an output at each step which is not carried forward to the remaining steps, but instead is returned from the overall scan function as an array of the $y_k$ values from each step.
We will not make use of this output, instead using the final carry as the result of the scan, and as such will ignore it.

To implement the forward-pass for $\varphi(s)$ in Eq. \eqref{eqn_sm:mps_ti_def}, the inputs $x_{k}$ are vectors defined as $x_{k} = (1 - s_{k}, s_{k})$ with $k = 1, 2, ..., L$ i.e. $L=T$. The carry is initiated as the identity matrix, $C = \mathcal{I}$, with shape $(\chi, \chi)$. The carry output of $f$ is then defined in terms of the tensor components $A_{s, \alpha, \beta} = [A_{s}]_{\alpha, \beta}$ as,
\begin{align}
    C_{\alpha, \gamma} &= \sum_{\beta = 0}^{\chi -1} \sum_{n = 0}^{1} C_{\alpha, \beta} A_{n, \beta, \gamma} [x_{k}]_{n} ~.
\end{align}
The additional output $y_{k}$ is not required and can be discarded. Collecting the inputs and outputs as vectors, $x$ and $y$, then the value of $\varphi(s)$ is then given by applying the scan-function for this $f$, followed by a trace over the carry,
\begin{align}
C, y &= \mathrm{scan}_{f}\left(C, x\right) \nonumber \\
    \varphi(s) &= \Tr\left[C \right] ~.
\label{eqn_sm:vf_scan}
\end{align}

Implemented in this manner, e.g. as found in \cite{SM_Code}, the gradient of the value-function can be easily evaluated using auto-differentiation.

\subsection{Policy Function Approximations For Local Kinetic Constraints}

Before turning to details of implementing the forward passes for $\pfa$ in the east model and ASEP, we briefly discuss here how the constraints on the dynamics can be captured explicitly in the structure of $\pfa$ via the constraint function $\mathcal{C}(a,s)$. To this end, recall first that in the main text we defined the policy function approximation as,
\begin{align}
\pfa = \mathcal{C}(a, s) \frac{\varphi(a,s)^{2}}{\mathcal{N}(s)} ~ ,
\label{eqn_sm:policy_defn}
\end{align}
where 
\begin{align}
    \varphi(a, s) &= \Tr\left[A_{a_{1}s_{1}} A_{a_{2}s_{2}} ... A_{a_{L}s_{L}}\right] ~.
\label{eqn_sm:mpo_ti_def}
\end{align}
Here, $\mathcal{N}(s) = \sum_{a | \mathcal{C}(a, s) = 1} \varphi(a,s)^{2} = \sum_{a} \mathcal{C}(a,s) \varphi(a,s)^{2}$ and $\mathcal{C}(a, s)$ is the constraint function, which returns one if an action $a$ is possible given a state $s$, or zero otherwise.

The constraint function, $\mathcal{C}(a,s)$, allows us to include explicit constraints on the actions selected by our function approximations. This is particularly powerful when modelling the dynamics of spin systems whose constraints are both known and such that only a few states can be reached from any other. In that case we can construct $\mathcal{C}(a, s)$ so that our policy reflects these constraints exactly, whereas in other scenarios this must be approximated or learnt.

\textbf{Single Spin-Flip Constraint:} In the dynamics studied in the main-text, we consider two varieties of constraint. The first, which pertains to both models, requires that at most a single spin is allowed to flip at a given time. For convenience, we will define the set of actions $\tilde{a}_{k}$ for $k \in [1,L+1]$ that represent a flip at site $k$ when $k \le L$ and no flip at any site when $k = L + 1$ (in which case the state is unchanged by the action). Taking $L=4$ for illustration, in terms of the variables in the main text where $a = (a_{1}, a_{2}, a_{3}, a_{4})$ with each $a_{k}$ indicating a flip at site $k$, we have $\tilde{a}_{1} = (1,0,0,0)$, $\tilde{a}_{2} = (0,1,0,0)$, $\tilde{a}_{3} = (0,0,1,0)$, $\tilde{a}_{4} = (0,0,0,1)$, and $\tilde{a}_{5} = (0,0,0,0)$. This constraint can be included in $\pfa$ straightforwardly via the choice,
\begin{align}
    \mathcal{C}(a,s) = \sum_{k=1}^{L+1} \delta_{\tilde{a}_{k}, a}.
    \label{eqn_sm:one_flip_constraint}
\end{align}

With the choice of constraint function \eqref{eqn_sm:one_flip_constraint}, sampling $\pfa$ will select actions only from the $L+1$ possibilities $\tilde{a}_{k}$, with other actions having strictly zero probability of occurring. As such, sampling can equivalently be performed by selecting an action from the probabilities $\lbrace \pi_w(\tilde{a}_{k}|s) \rbrace_{k=1}^{L+1}$ alone. This simplifies the problem of sampling $\pfa$, because the normalisation factor $\mathcal{N}(s)$ [c.f. \eqref{eqn_sm:policy_defn}] --which in general is hard to calculate for conditional probabilities-- can be calculated explicitly by enumerating $\mathcal{C}(\tilde{a}_{k}, s) \varphi(\tilde{a}_{k},s)^{2}$ for all $k = 1, 2, ..., L+1$. Thus, at most $L+1$ computations are required to sample an action from a policy with this constraint. While, in principle, these could be computed in parallel, for the problems here we present an alternative method (as applied in the main text) where the policy is instead sampled via a ``sweep", similar to that performed for more standard tensor network algorithms.  

\textbf{Local Kinetic Constraint:} The second aspect of the constraints is the local kinetic constraint. Here, whether a spin at site $k = 1, 2, ..., L$ can flip depends only on the states of the neighbouring sites at $k-1, k$ and $k+1$. For example, in the case where the possibility of $\tilde{a}_{k}$ depends on a three-site neighbourhood we can further write that,
\begin{align}
    \mathcal{C}(a,s) = \sum_{k=1}^{L} \delta_{\tilde{a}_{k}, a} \mathcal{C}(\tilde{a}_{k}, s_{k-1}, s_{k}, s_{k+1}) +  \delta_{\tilde{a}_{L+1}, a} \mathcal{C}(\tilde{a}_{L+1}, s) ~.
\end{align}
Note that here we have separated out the ``no-flip" action, $\tilde{a}_{L+1}$, as this must typically be treated separately in a given problem.

While both the east model and ASEP are subject to local kinetic constraints, the specific form of the local constraint function, $\mathcal{C}(\tilde{a}_{k}, s_{k-1}, s_{k}, s_{k+1})$, will depend on the model in hand. As such, the function approximations for the polices will differ slightly and, therefore, so will the implementation of the forward passes.

\subsection{Forward Pass for $\pfa$ in the East Model}

We now describe the implementation of the forward pass for $\pfa$ in the east model, see \cite{SM_Code} for an explicit example of implementation. In the east model, a spin can only flip if the spin to its left it up. As such, this local kinetic constraint can be captured by the local constraint function,
\begin{align}
    \mathcal{C}(\tilde{a}_{k}, s) = \mathcal{C}(\tilde{a}_{k}, s_{k-1}, s_{k}, s_{k+1}) = s_{k-1} ~ \text{for} ~ k \in [1, L] ~.
\end{align}
For the case of no-flips, $\tilde{a}_{L+1}$, we take this to be always possible unless every spin is up, i.e.,
\begin{align}
    \mathcal{C}(\tilde{a}_{L+1}, s) = 1 - \delta_{s, (1,1,1,...,1)} ~.
\end{align}

Due to the constraint Eq. \eqref{eqn_sm:one_flip_constraint}, we need consider only the possible actions $\tilde{a}_{k}$. For these actions, the matrix product operator used in the function approximation for the policy \eqref{eqn_sm:policy_defn} takes the form,
\begin{align}
\varphi(\tilde{a}_{k}, s) &= \Tr\left[ \left(\prod_{l=1}^{k-1} A_{0 s_{l}}\right) A_{1 s_{k}} \left(\prod_{l = k+1}^{L} A_{0 s_{l}}\right)\right] ~.
\end{align}

We then define the product of matrices $A_{0 s_{m}}$ to the left and right of some site $m$ as,
\begin{align}
    M^{\text{left}}_{m} = \prod_{l = 1}^{m-1} A_{0 s_{l}} ~,
\end{align}
and
\begin{align}
    M^{\text{right}}_{m} = \prod_{l = m + 1}^{L} A_{0 s_{l}} ~.
\end{align}
These matrices can be constructed iteratively as,
\begin{align}
    M^{\text{left}}_{m+1} = M^{\text{left}}_{m} A_{0 s_{m}} ~,
    \label{eqn_sm:left_mat_em}
\end{align}
and
\begin{align}
    M^{\text{right}}_{m-1} = A_{0 s_{m}} M^{\text{right}}_{m}  ~.
    \label{eqn_sm:right_mat_em}
\end{align}
To relate these to the probabilities of taking a given action, we then define the left-environment,
\begin{align}
    \mathcal{E}^{\text{left}}_{m} = \mathcal{C}\left(a, s\right) M^{\text{left}}_{m} A_{1 s_{m}}~,
\end{align}
which includes the site $m$ and the constraint function, $\mathcal{C}\left(a, s\right)$.  With this,
\begin{align}
    \varphi(\tilde{a}_{k}, s) &= \Tr\left[ \mathcal{E}^{\text{left}}_{k}  M^{\text{right}}_{k}\right] ~.
    \label{eqn_sm:factors_em}
\end{align}

The iterative form of Eq. \eqref{eqn_sm:left_mat_em} and the expression Eq. \eqref{eqn_sm:factors_em} shows that we can construct all required $\varphi(\tilde{a}_{k}, s)$ iteratively by ``sweeping" from left to right (obtaining $\mathcal{E}^{\text{left}}_{k}$ for $k=1,2,3,...,L)$ and then from right to left (obtaining $M^{\text{right}}_{k}$ for $k = L, L-1, ..., 1$). These sweeps can be implemented efficiently with the use of a \text{scan} function, just as with $\vfa$ [c.f. Eq. \eqref{eqn_sm:vf_scan}]. However, while for $\vfa$ just one sweep to the right was required (i.e. a single $\text{scan}$ function), here an additional sweep to the left is required.

Specifically, the rightward sweep is implemented by a scan of the ``right-step" function $f_{R}$, such that $C, y_{k} = f_{R}(C, x_{k})$. The carry in this case consists of two objects, $C = \lbrace \mathcal{E}_{L}, M \rbrace$. The first is the set of left environments, $\mathcal{E}_{L}$, whose components are the left-environments for a specific site $m$, i.e. $[\mathcal{E}^{\text{left}}]_{m} = \mathcal{E}_{m}^{\text{left}}$. The second object is the matrix $M$, which at the start of any step $k$ is simply equal to $M_{k}^{\text{left}}$. At step $k$, these are updated as,
\begin{align}
    [\mathcal{E}_{L}]_{k} \to \mathcal{C}(\tilde{a}_{k}, s) M A_{1 s_{k}} ~, \\
    M \to M A_{0 s_{k}} ~. 
\end{align}
As with $\vfa$, the output $y_{k}$ of $f_{R}$ is not needed and can be discarded. Scanning $f_{R}$ then gives,
\begin{align}
    \lbrace \mathcal{E}_{L}, M \rbrace, y = \text{scan}_{f_{R}}(\lbrace \mathcal{E}_{L}, M \rbrace, x) ~,
    \label{eqn_sm:scan_to_right_em}
\end{align}
where once again the input is a set of vectors, $x = \lbrace (1 - s_{l}, s_{l}) \rbrace_{l=1}^{L}$.

According to Eq. \eqref{eqn_sm:factors_em}, to complete the computations of $\varphi(\tilde{a}_{k}, s)$ we must then compute $M_{k}^{\text{right}} ~ \forall ~ k \in [1, L]$. From the iterative expression Eq. \eqref{eqn_sm:right_mat_em} it is clear how to achieve this, see \cite{SM_Code} for an example implementation.

\subsection{Forward Pass for $\pfa$ in the ASEP}

In the ASEP we consider, the local kinetic constraint is such that particles (i.e. up spins) can move to the left/right only if there is an unoccupied space (i.e. a down spin) in that position. Moreover, analogous to the east model where at most only a single spin could flip per time-step, for the ASEP at most a single particle can move left or right. These constraints can be realised in $\pfa$ by setting the constraint function $\mathcal{C}(a, s)$ such that the only possible actions are the $L$ actions that flip pairs of spins, $\tilde{a}_{k}: (s_{k}, s_{k+1}) \to (1 - s_{k}, 1 - s_{k+1})$, and the no-flip action $\tilde{a}_{L+1}$. Taking $L=4$ for illustration, in terms of the fundamental single-site spin flip actions: $\tilde{a}_{1} = (1, 1, 0, 0)$, $\tilde{a}_{2} = (0, 1, 1, 0)$, $\tilde{a}_{3} = (0, 0, 1, 1)$, $\tilde{a}_{4} = (1, 0, 0, 1)$ and $\tilde{a}_{5} = (0, 0, 0, 0)$. The constraint function is then,
\begin{align}
    \mathcal{C}(a, s) &= \sum_{k=1}^{L} \delta_{a, \tilde{a}_{k}} \mathcal{C}(\tilde{a}_{k}, s) + \delta_{a, \tilde{a}_{L+1}} \mathcal{C}(\tilde{a}_{L+1}, s) ~, \\
    &= \sum_{k=1}^{L} \delta_{a, \tilde{a}_{k}} \mathcal{C}(\tilde{a}_{k}, s_{k}, s_{k+1}) + \delta_{a, \tilde{a}_{L+1}} \mathcal{C}(\tilde{a}_{L+1}, s) ~, \\
    &= \sum_{k=1}^{L} \delta_{a, \tilde{a}_{k}} \left[(1 - s_{k})s_{k+1} + s_{k}(1-s_{k+1}) \right] + \delta_{a, \tilde{a}_{L+1}} \mathcal{C}(\tilde{a}_{L+1}, s) ~.
\end{align}
Here, in the third line, the choice of local constraint function, $\mathcal{C}(\tilde{a}_{k}, s_{k}, s_{k+1})$, ensures action $\tilde{a}_{k}$ only occurs when there is a single particle in the pair of sites to be flipped.

The special case of no-flip, realised in $\mathcal{C}(\tilde{a}_{L+1}, s)$, can always occur except when no particles have a neighbour (e.g. $(0,0,1,0)$ for $L=4$). This can be expressed as,
\begin{align}
    \mathcal{C}(\tilde{a}_{L+1}, s) = 1 - \mathds{I}\left(\sum_{k=1}^{L} s_{k} s_{k+1} = 0\right) ~,
\end{align}
where we have introduced the so-called indicator function, $\mathds{I}$, which returns $1$ or $0$ if the condition in its argument is true or false respectively.

The expression for $\varphi(\tilde{a}_{k}|s_{t})$ in the case of the ASEP is,
\begin{align}
\varphi(\tilde{a}_{k}, s) &= \Tr\left[ \left(\prod_{l=1}^{k-1} A_{0 s_{l}}\right) A_{1 s_{k}} A_{1 s_{k+1}}  \left(\prod_{l = k+2}^{L} A_{0 s_{l}}\right)\right] ~ .
\label{eqn_sm:factor_asep_ak}
\end{align}
Using the previous definitions for the left-environments, $\mathcal{E}^{\text{left}}_{k}$ and the matrices $M^{\text{right}}_{k}$, this is equivalent to
\begin{align}
    \varphi(\tilde{a}_{k}, s) &= \Tr\left[ \mathcal{E}^{\text{left}}_{k} A_{1 s_{k+1}}  M^{\text{right}}_{k+1}\right] ~.
    \label{eqn_sm:factors_asep}
\end{align}
Expressed in this manner, it is clear that to implement the forward pass for $\varphi(\tilde{a}_{k}, s)$ we can again apply a $\text{scan}$ function, just as with the east model. Indeed, the details of this procedure are very similar to those of the east model outlined previously, although slightly more involved due to the fact the $L+1$ potential actions $\tilde{a}_{k}$ change two sites rather than one. For further details, a full example implementation is given in \cite{SM_Code}.

\end{document}